\documentclass[11pt]{article}
\begin{document}

\title{APS/DFD Gallery of Fluid Motion 2013 (\#102302)\\
Why does a beer bottle foam up\\ after a sudden impact on its mouth?}

\author{Javier Rodr\'{\i}guez-Rodr\'{\i}guez$^{1,} $\footnote{e-mail: javier.rodriguez@uc3m.es}, Almudena Casado-Chac\'on$^1$ and Daniel Fuster$^2$\vspace{5mm}\\
{\small $^1$ Fluid Mechanics Group, Carlos III University of Madrid, SPAIN}\\
{\small $^2$ CNRS (UMR 7190), Universit\'e Pierre et Marie Curie, Institut Jean le Rond d'Alembert, FRANCE}
}%

\date{}

%
%

\maketitle

Besides the obvious interest in the field of recreative physics -which initially attracted our attention to the problem- understanding the formation of foam in a supersaturated carbonated liquid after an impact on the container involves the careful physical description of a number of processes of great interest in bubble dynamics and multiphase flows in general. In order of appearance in this problem: strong pressure wave propagation in bubbly liquids, bubble-bubble interaction in clusters, bubble collapse and break-up, diffusive bubble-liquid mass transfer and the dynamics of bubble-laden plumes. As a matter of fact, very similar physical phenomena are found in the study of oil reservoirs using seismic waves or in the formation of mud volcanoes.\\

To understand the problem at hand, the foaming-up of the beer, experiments have been conducted impacting commercial beer bottles (Mahou 25 cl) with different brass weights dropped from variable heights. Two sets of experiments were carried out: firstly, bottles were filled with deionized water and a hydrophone (Reson TC-4035) was placed at different depths to quantify the pressure wave propagation in the liquid. In a second set of experiments, the bottles contained beer and the evolution of the bubbles after the impact was recorded with a high-speed camera NAC HX-3.\\

In few sentences, the whole chain of processes occurring in the bottle after the impact can be summarized as follows: a sudden vertical impact on the mouth of a beer bottle generates a compression wave that propagates through the glass towards the bottom. When this wave reaches the base of the bottle, it is transmitted to the liquid as an expansion wave that travels to free surface, where it bounces back as a compression wave. The peak pressure of the different waves can be estimated using the classical theory of impact on solids \cite{Goldsmith2001}. Waves bounce back and forth several more times until they are damped out. This train of expansion-compression waves drives the forced cavitation of existing air pockets, similar to those found in other carbonated beberages \cite{Liger-Belair_etalJPCB2005}, leading to their violent collapse. Notice that, if only the first expansion wave existed, bubbles would just oscillate around their new equilibrium radius without collapsing. So the presence of a relatively close free surface is key to form the train of expansion-compression waves that drive the bubble implosions.\\

Clouds of very small daughter bubbles are generated upon these collapses, that expand much faster than their mothers due to their smaller size and thus increased surface to volume ratio \cite{EpsteinPlessetJCP50}. These rapidly growing bubble clusters effectively act as buoyancy sources, what leads to the formation of bubble-laden plumes whose void fraction increases quickly by several orders of magnitude, eventually turning most of the liquid into foam. Preliminary numerical simulations performed using Gerris Flow Solver \cite{Tomar_etalCF2010, PopinetJCP2003} confirm this point.\\

All the above processes are illustrated in the accompanying fluid dynamics video using our experimental results. A more detailed theoretical analysis of these processes will be given in talk A11.00006, and in a paper in preparation.\\

We acknowledge the support of the Spanish Ministry of Economy and Competitiveness through grant \#DPI2011-28356-C03-02.\\

\end{document}